\documentclass[aps,preprint,preprintnumbers,nofootinbib,showpacs]{revtex4}
\usepackage{graphicx,color,amsmath,amsxtra}

\newcommand{\newc}{\newcommand}

\newc{\be}{\begin{equation}}
\newc{\ee}{\end{equation}}
\newc{\beq}{\begin{eqnarray}}
\newc{\eeq}{\end{eqnarray}}

\begin{document}
\title{Hypermagnetic baryogenesis}
\author{
Kazuharu Bamba$^{1}$, C. Q. Geng$^{2}$ and S. H. Ho$^{2}$
}
\affiliation{
$^{1}$Department of Physics, Kinki University, Higashi-Osaka 577-8502, Japan\\
$^{2}$Department of Physics, National Tsing Hua University,
Hsinchu, Taiwan 300}
%\date{\today}

\begin{abstract}
We study a new scenario for baryogenesis due to the spontaneous breaking of
the $CPT$ invariance through the interaction between a baryon current and
a hypermagnetic helicity. The hypermagnetic helicity  (Chern-Simons number) of $U(1)_Y$
provides a $CPT$ violation background for the generation of baryons via sphaleron processes,
which protects these baryons from the sphaleron wash-out  effect in thermal equilibrium.
It is shown that if the present amplitude of the resultant magnetic fields are sufficiently large,
for a wide range mass scale (from  $\mathrm{TeV}$  to the Planck scale),
the observational magnitude of the baryon asymmetry of the Universe can be realized.
\end{abstract}

\preprint{KU-TP\ 016}
\maketitle
%====================================================================

The origin of the matter-antimatter asymmetry of the Universe is still an
unsolved problem. The magnitude of the baryon asymmetry of the Universe (BAU)
is characterized by the ratio of the baryonic number density $n_B$ to
the entropy density $s$, which is observationally estimated as
\beq \label{ratio}
\frac{n_B}{s} = 0.92_{\,-0.04}^{\,+0.06} \times 10^{-10}\,,
\eeq
by using the first year Wilkinson Microwave Anisotropy Probe (WMAP) data on
the anisotropy of the cosmic microwave background (CMB) radiation \cite{Spergel:2003cb}.
There exist various scenarios to explain the observational value in Eq. (\ref{ratio})
from the baryon symmetric universe
\cite{Dolgov:1991fr, Funakubo:1996dw,Dine:2003ax}.
Under the assumption of the $CPT$ invariance, Sakharov stated that three
conditions are necessary to generate the BAU: (i) baryon number violation,
(ii) $C$ and $CP$ violation, (iii) a departure from thermal equilibrium
\cite{Sakharov:1967dj}. However, if the $CPT$ invariance is
violated in the early universe, the condition (iii) is no longer  necessary
\cite{Dolgov:1991fr,Bertolami:1996cq}. An effective
mechanism of this idea with a derivative scalar field coupled to the baryon current was first proposed
 by Cohen and Kaplan
\cite{Cohen:1987vi}, which is called ``spontaneous baryogenesis".
If the time derivative of the scalar field has a
non-zero expectation value, this interaction violates the $CPT$
invariance spontaneously and hence an effective chemical potential difference between
baryons and antibaryons is produced.

On the other hand, it has been pointed out
\cite{Joyce:1997uy, Giovannini:1997gp, Thompson:1998qz, Vachaspati:1994ng}
that hypercharge electromagnetic fields can play a significant role in the
electroweak (EW) scenario
\cite{Cohen:1993nk, Rubakov:1996vz, Funakubo:1996dw}
for baryogenesis.
In particular, Giovannini and Shaposhnikov (GS) \cite{Giovannini:1997gp} have shown that
the Chern-Simons number stored in the hypercharge electromagnetic fields,
corresponding to the hypermagnetic helicity, is converted into fermions at
the electroweak phase transition (EWPT) due to the anomaly if  it is strongly first order
\cite{Giovannini:1997gp},
while the hypermagnetic fields are replaced by the ordinary magnetic fields, which survive
after the EWPT up to the present time and hence can be cosmic magnetic fields observed in galaxies
and clusters of galaxies.

The most natural origin of large-scale hypermagnetic fields before
the EWPT is hypercharge electromagnetic quantum fluctuations
in the inflationary stage \cite{Turner:1987bw, mag review}.
If the conformal invariance of the Maxwell theory is broken by some mechanism
in the inflationary stage
\cite{Turner:1987bw, Ratra:1991bn},
%, Bamba-mag1,Martin:2007ue,scalar-string, Garretson:1992vt, Giovannini-mag, Field:1998hi, Bamba-mag2,
%Dolgov:1993vg},
hypercharge electromagnetic quantum fluctuations exist
even in the conformal flat Friedmann-Robertson-Walker (FRW) spacetime.
Furthermore, if the hypercharge electromagnetic fields
couple to an axion-like pseudoscalar field
with a time-dependent expectation value, the hypermagnetic helicity can be generated
\cite{Brustein-hel, Giovannini-hel, Campanelli:2005ye, Bamba:2006km}.
(Incidentally, in Ref.~\cite{Brustein-hel,Guendelman:1991se} baryogenesis due to the above coupling has been discussed. 
Moreover, in Ref.~\cite{Copi:2008he} helical magnetic fields from sphaleron 
decay and baryogenesis have recently been considered.) 
In this case, the scale of the hypermagnetic fields with the helicity can be
larger than or equal to the Hubble horizon.
As a result, the homogeneous baryogenesis over the whole
present universe can be realized \cite{Bamba:2006km}.

In this Letter, we propose a new scenario for baryogenesis
 through the $CPT$-even dimension-six Chern-Simons-like
interaction given by Geng, Ho and Ng (GHN) in Ref. \cite{Geng:2007va}.
 We will concentrate on the interaction between a baryonic current
 and a hypermagnetic helicity. This type of the helicity can be produced much
before the EWPT as hypercharge electromagnetic quantum fluctuations in the
inflationary stage through some breaking mechanism of the conformal invariance
of the hypercharge electromagnetic field.
It is clear that, in the standard model (SM) we cannot use the GS mechanism \cite{Giovannini:1997gp} to
induce baryogenesis  as the EWPT is not first order \cite{Kajantie:1996mn} 
and the resultant baryons will be destroyed by the sphaleron processes
\cite{Kuzmin:1985mm} (see also \cite{Manton-sph}).
In our new scenario, however, because the $CPT$ invariance is broken
spontaneously \cite{Kostelecky} by the hypermagnetic helicity with its non-zero classical expectation value,
the resultant baryons will not be destroyed by the sphaleron processes
even in the SM  \cite{Carroll:2005dj}.

In our study, we will adopt the Heaviside-Lorentz units and
$k_\mathrm{B} = c = \hbar = 1$ and assume the spatially flat FRW spacetime with the metric
\beq
ds^2&=&-dt^2 + a^2(t)d{\vec{x}}^2\,,
\eeq
where $a(t)$ is the scale factor.

We start with
the $CPT$-even dimension-six Chern-Simons-like effective interaction \cite{Geng:2007va}:
\beq \cal{L}_{\mathrm{CS}}&=&
-{\beta\over
M^{2}}j_{\mu}(\emph{Y}_{\nu}\emph{\~{Y}}^{\,\,\mu\nu}+\partial_{\nu}S^{\mu\nu})\,,
\label{Lcs}
\eeq
where
$\emph{Y}_{\nu}$ is the $U(1)_Y$ gauge field,
$\emph{\~{Y}}^{\,\,\mu\nu} \equiv \frac{1}{2}\epsilon^{\mu\nu\rho\sigma}\emph{Y}_{\rho\sigma}$
is the dual of the $U(1)_Y$ hypercharge field strength tensor,
$\emph{Y}_{\mu\nu}=\partial_{\mu}Y_{\nu}-\partial_{\nu}Y_{\mu}$,
 $j_{\nu}$ is a fermion current, $\beta$ is a dimensionless coupling parameter, 
 %%%%%%%%%%
 %\textcolor {red}{
 $S$ is the St$\ddot{u}$ckelberg for maintaining the general gauge invariance \cite{jackiw},
 %} 
 %%%%%%%%%%
 and $M=\Lambda/4\pi$ with $\Lambda$ being the scale of the effective interaction.
Here, $\epsilon^{\mu\nu\rho\sigma}$ is the totally anti-symmetric
Levi-Civita tensor with the normalization of $\epsilon^{0123}=+1$.
Note that in Eq. (\ref{Lcs}) we have extended  the electromagnetic field and  neutrino current
 in  the interaction given by GHN \cite{Geng:2007va} to the hypercharge field and any fermion current, respectively.

%%%%%%%%%%%%
In Ref.~\cite{Geng:2007va}, it is concluded that the fermion current $j_{\nu}$
to a comoving observer has the form:
\beq \label{current}
\emph{j}_{\mu}= \bar{\psi}\gamma_{\mu}\psi =
(n_{\psi}-n_{\bar{\psi}} , \vec{0}),
\eeq
where $n_{\psi}$ and $n_{\bar{\psi}}$ are the number densities of
the fermion $\psi$ and antifermion $\bar{\psi}$, respectively.
%%%%%%%%%%%
%\textcolor {red}{
It is interesting to note that, as pointed out in Ref. \cite{Geng:2007va}, the modified interaction would originate from superstring theory, in which the role of the St$\ddot{u}$ckelberg field is
played by the anti-symmetric Kalb-Ramond field $B_{\mu\nu}$. 
In the homogeneous and isotropic universe, it is reasonable to assume that this $B_{\mu\nu}$ field is only a function of the cosmic time $t$ \cite{DiGrezia:2004md}. Then the second term in Eq. (\ref{Lcs}) becomes $j_{\mu}\epsilon^{\mu\nu\rho\sigma} \partial_{\nu} B_{\rho\sigma}$, which vanishes in the spatially flat FRW spacetime.
% }
%%%%%%%%%%%
Hence, the Lagrangian in Eq.\ (\ref{Lcs}) reduces to
\beq
\cal{L}_{\mathrm{CS}}&=&
-{\beta\over
M^{2}}j_{0}\ \vec{Y} \cdot (\nabla\times\vec{Y}),
\label{Leff}
\eeq
where  $j_0=n_{\psi}-n_{\bar{\psi}}$.
%We remark that Eq. (\ref{Leff}) is gauge invariance \cite{Geng:2007va}, whereas
%${\cal L}_{\mathrm{CS}}$ of Eq. (\ref{Lcs}) may be not generally.
 %To formally maintain  the  gauge invariance in Eq. (\ref{Lcs}),
%we can introduce the St$\ddot{u}$ckelberg field \cite{Geng:2007va,jackiw}. 
%As pointed out in Ref. \cite{Geng:2007va}, the modified interaction would originate
 %from superstring theory.

We now consider that $j_{\mu}$ is the baryon current and there exists a non-vanishing
hypermagnetic helicity before the EWPT.
The interaction between the baryon current and the hypermagnetic helicity
in Eq.\ (\ref{Leff}) splits the spectrum of the baryons and antibaryons by
giving them  effective chemical potentials,
\beq
\mu_{B}=-\mu_{\bar{B}}\equiv\mu=
{\beta\over M^{2}}
\left< \vec{Y} \cdot (\nabla\times\vec{Y}) \right>,
\label{chemical}
\eeq
which lead to the net baryonic number density in
the thermal equilibrium as \cite{Kolb and Turner}
\beq
n_B \equiv n_b-n_{\bar{b}}=
\frac{g_b T^3}{6} \left( \frac{\mu}{T} \right) + \emph{O}
\left( \frac{\mu}{T} \right)^3,
\label{asymmetry}
\eeq
where $n_b$ and $n_{\bar{b}}$ are the baryonic and antibaryonic number
densities, respectively,
$g_b$ counts the internal degrees of freedom of the baryons, and $T$ is the
background temperature of the Universe.

The density of the hypermagnetic helicity is defined by
\beq
h_B \equiv \vec{Y} \cdot (\nabla\times\vec{Y})
= \vec{Y} \cdot {\vec{H}}_{Y}\,,
\eeq
where ${\vec{H}}_{Y}$ is the hypermagnetic field. The energy density of
the hypermagnetic fields,
${\rho}_{H_{Y}} \equiv
| \vec{H}_{Y}|^2/2$,
and the density of the hypermagnetic helicity have to satisfy the
realizability condition \cite{Campanelli:2005ye, Campanelli:2005jw}:
\beq \label{condition}
h_B \leq 2L {\rho}_{H_{Y}}\,,
\eeq
where $L$ is the coherence scale of the hypermagnetic fields.
In the case of the hypermagnetic fields with its maximum helicity,
the effective chemical potential is given by
\beq
\mu
&=& {\beta\over M^{2}} <h_B>  \nonumber  \\[2.5mm]
&=& {\beta\over M^{2}} \left( 2L{\rho}_{H_{Y}} \right)
= {\beta\over M^{2}} L | \vec{H}_{Y} |^2\,.
\label{effective-chemical-potential}
\eeq
After the freeze-out temperature $T_{\mathrm{f}}$, it follows from
Eq.\ (\ref{asymmetry}) that the density of the residual baryonic number is
given by
\beq
n_B(T_{\mathrm{f}})=\frac{g_b T_{\mathrm{f}}^3}{6}
\left[ \frac{\mu \left(T_{\mathrm{f}} \right)}{T_{\mathrm{f}}} \right]\,,
\label{freeze-out}
\eeq
where we have neglected the term of $\emph{O} \left( \mu/T \right)^3$.
After reheating following inflation, a number of charged particles are
produced, so that the conductivity of the Universe
is much larger than the Hubble parameter at that time.
The hypermagnetic fields evolve
as $| {\vec{H}}_{Y}| \propto a^{-2}\propto g_s^{2/3} T^2$
due to the magnetic flux conservation, while the entropy
density and the coherence scale of the hypermagnetic fields behave as
$s\propto g_sT^3$ and
$L \propto a \propto g_s^{-1/3}T^{-1}$, respectively, where $g_s$
represents  the total number of effective massless degrees of freedom referring
to the entropy density of the Universe \cite{Kolb and Turner}.
Note that reheating occurs much before the EWPT.
We emphasize that in our scenario, since the sphaleron effect is served as the source of the
baryon number violation,    the freeze-out temperature $T_{\mathrm{f}}$ corresponds
to the background temperature at the EWPT to be $T_{\mathrm{EW}}\sim\; 150$ GeV.

Consequently,
after putting in the corrected baryon numbers  and  three generations of quarks,
it follows from Eqs.~(\ref{effective-chemical-potential}) and (\ref{freeze-out})
 that the baryon-to-entropy ratio at the
freeze-out temperature $T_{\mathrm{f}}$ is expressed as
\beq
\frac{n_B}{s}(T_{\mathrm{f}})
&=& \beta g_b
\left( \frac{T_{\mathrm{f}}}{M} \right)^2 \frac{L_0{B_0}^2}{s_0}
\label{expression-of-ratio} \\[2.5mm]
&\approx&
%1.9 *3 *2
1.2 \times 10^{40}
\beta \left( \frac{T_{\mathrm{f}}}{M} \right)^2
\left( \frac{B_0}{[\mathrm{G}]} \right)^2\,,
\label{asymmetry-ratio}
\eeq
where the subscript suffix `$0$' represents the quantities at the present
time, and $B_0 \equiv | \vec{B}_0| $ is the present field strength
of the magnetic fields.
Note that in Eq.\ (\ref{expression-of-ratio}), we have rescaled
the coherence scale of the magnetic fields,
the amplitude of the magnetic fields, and the entropy density from
the values at $T_{\mathrm{f}}$ to the values at the present time, respectively.
% by using $L\propto a \propto g_s^{-1/3}T^{-1}$,
%$B \propto a^{-2} \propto g_s^{2/3} T^2$, and $s\propto g_sT^3$.
Moreover, in Eq.\ (\ref{asymmetry-ratio}), we have used that $g_b=2$ and
 the coherent length of the magnetic fields at the present
time $L_0$ is equal to the current horizon scale $H_0^{-1}$ because we are
considering the homogeneous baryogenesis over the whole present universe.

As an illustration, by taking $\beta \sim 1$,
the freeze-out temperature
$T_{\mathrm{f}} = T_{\mathrm{EW}} \sim 150 \, \mathrm{GeV}$,
and the present field strength of the magnetic fields on the horizon scale
$B_0 \sim 10^{-9} \mathrm{G}$, which is the upper limit on the present field
strength of the primordial magnetic fields on the horizon scale obtained by
carrying out a statistical analysis for the angular anisotropy of the CMB
radiation \cite{Barrow:1997mj}, we find from Eq.\ (\ref{asymmetry-ratio}) that
the resultant value of the baryon-to-entropy ratio, $n_B/s \sim 10^{-10}$, can be realized
if the mass scale is
$\Lambda=4\pi M \sim M_{Planck}$.
Moreover, for $\beta \sim 1$,
$T_{\mathrm{f}} = T_{\mathrm{EW}} \sim 150 \, \mathrm{GeV}$,
and $B_0 \sim 10^{-24} \mathrm{G}$,
it follows from Eq.\ (\ref{asymmetry-ratio}) that  $n_B/s \sim 10^{-10}$ can be obtained
with $\Lambda \sim 1\mathrm{TeV}$.

%%%%%%%%%%%%%%%
%%% Summary %%%
%%%%%%%%%%%%%%%
In summary, we have proposed a new scenario for baryogenesis due to the
spontaneous breaking of the $CPT$ invariance through the interaction
in Eq.\ (\ref{Lcs}). The hypermagnetic helicity
can be generated much before the EWPT as hypercharge
electromagnetic quantum fluctuations in the inflationary stage through some
breaking mechanism of the conformal invariance of the hypercharge
electromagnetic field.
In this scenario, the resultant baryons will not be destroyed by
the sphaleron processes even if the EWPT is not  first order
due to the spontaneous breaking of the $CPT$ invariance.
We have found that if there are magnetic fields with the field strength
$B_0$ being  $10^{-24}\,$-$\,10^{-9} \mathrm{G}$ on the horizon scale at the
present time, while the corresponding
mass scale $\Lambda$ in terms of a baryon current interacting to a hypermagnetic
helicity is $\mathrm{TeV}\,$-$\, M_{Planck}$, the resultant value of the baryon-to-entropy ratio,
$n_B/s \sim 10^{-10}$, can be achieved, which is consistent with the magnitude
of the BAU suggested by observations obtained from the WMAP.
%%%%% Remark %%%%%%
Finally, we remark that if the effective
Chern-Simons-like
interaction in Eq. (\ref{Lcs}) is originated from superstring theory with  $\Lambda\sim M_{Planck}$,
$B_0$ should be  $\geq 10^{-10} \mathrm{G}$,
which can be tested \cite{Caprini:2003vc, Kahniashvili:2006zs}
in future experiments such as PLANCK \cite{Planck}, SPIDERS
(post-PLANCK) \cite{SPIDERS} and Inflation Probe (CMBPol mission) in the Beyond Einstein program of
NASA \cite{CMBPol}.\\

%the following point:
%According to Refs.~\cite{Caprini:2003vc, Kahniashvili:2005xe,Kahniashvili:2006zs},
%if the present field strength of the primordial magnetic fields is larger than or equal to $10^{-10} \mathrm{G}$,
%and if its spectrum is nearly scale invariant and its helicity is close to the
%maximum, such a primordial magnetic field can be detected by the future
%measurements of parity-odd cross correlations between temperature and
%$B$-polarization anisotropies of the CMB radiation, and between
%$E$- and $B$-polarization anisotropies of it, which could be obtained
%from post-PLANCK experiments such as SPIDERS
%\footnote{http://www.astro.caltech.edu$/\sptilde\mathrm{lgg}/{\mathrm{spider}}_{\_}$front.htm}
%and Inflation Probe (CMBPol mission) in the Beyond Einstein program of
%NASA\footnote{http://universe.nasa.gov/program/probes/inflation.html}
%\cite{Saito:2007kt}.
%Such a primordial magnetic field with the field strength
%$\geq 10^{-10} \mathrm{G}$ and a nearly scale invariant spectrum
%can be generated from electromagnetic quantum fluctuations due to the
%breaking of the conformal invariance of the Maxwell
%theory during inflation \cite{Ratra:1991bn, Bamba-mag1}.
%Hence, if the primordial helical magnetic fields with the field strength
%$\sim 10^{-9} \mathrm{G}$ are detected by some future observations of
%the polarization of the CMB radiation, from Eq.\ (\ref{asymmetry-ratio})
%we can determine the mass scale as
%$M \sim 10^{18} \mathrm{GeV}$, i.e., about the Planck scale.
%%%%%%%%%%%%%%%

%%%%%%%%%%%%%%%%%%%%%%%
%%% Acknowledgments %%%
%%%%%%%%%%%%%%%%%%%%%%%
\noindent
{\bf Acknowledgments}

K.B. thanks all the members of the particle physics
group at National Tsing Hua University for their very kind hospitality.
This work is supported in part by
the open research center project
at Kinki University (K.B.)
and the National Science Council of
R.O.C. under Grant \#:
NSC-95-2112-M-007-059-MY3 (C.Q.G and S.H.H).
%%%%%%%%%%%%%%%%%%%%%%%

\end{document}